\documentclass[reprint,
amsmath,amssymb,
aps]{revtex4-2}
\usepackage{graphicx}

\begin{document}

\title{All Optical Integrated MOEMS Optical Coherence Tomography System}

\author{Ashish Singh Bais}
\affiliation{Applied Photonics Laboratory, Department of Applied Physics and Optoelectronics,
Shri G. S. Institute of Technology and Science, 23 Park Road Indore, MP 452003 India.}
\author{Lokendra Singh Chouhan}
\affiliation{Applied Photonics Laboratory, Department of Applied Physics and Optoelectronics,
Shri G. S. Institute of Technology and Science, 23 Park Road Indore, MP 452003 India.}
\altaffiliation{Also at St. Paul HSS, Indore, MP 452001  India}
 \author{Joseph Thomas Andrews}
\email{Corresponding author jtandrews@sgsits.ac.in}
\affiliation{Applied Photonics Laboratory, Department of Applied Physics and Optoelectronics,
Shri G. S. Institute of Technology and Science, 23 Park Road Indore, MP 452003 India.}

\begin{abstract}
Integrating all optical components of an optical coherence tomography (OCT) device into a single chip is a non-trivial and a challenging job. The design and development of such a
lab-on-a chip will be possible only via Micro-Opto-Electro-Mechanical System (MOEMS) technology. The reproducible and integrated optical device fabrication would reduce cost and size many fold as compared to bulk or fiber optic OCT system. A
miniaturized OCT of size less than 5mm$^2$ area is
designed, simulated and optimized. The successful
fabrication of this device would help in point-of-contact
devices as well as embedded biomedical sensor
applications. Also, the design promises the possibility of fabrication of all optical components of OCT integrated into a single chip.
\end{abstract}
\maketitle

\section{Introduction}
Significant advantage of integrated optical devices are its
size, portability, mobility, multi-functional measurement capabilities, etc.,
over the bulk optical devices. Common methods
adopted for the integration multiple optical components
into a single chip is e-beam methods, pulse laser
deposition, epitaxy methods, chemical vapor deposition
or a cheaper photo-lithography.
The developed micro-opto-electro-mechanical systems (MOEMS),
once optimized can be used in industry, clinics,
point-of-contact devices, mobile devices, etc. \cite{Li}.

\begin{figure}
\begin{center}
\includegraphics[width=0.5\columnwidth]{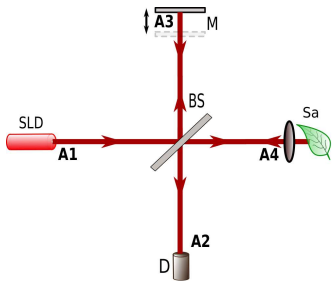}
\end{center}
\caption{\label{fig1} Schematic of an optical coherence tomography setup. SLD-Light source, D-Light detector, BS-Beam splitter, Sa-Sample, M-Scanning mirror. A1-Source, A2-Detector, A3-Reference  and A4-Sample arms of an Michelson Interferometer.}
\end{figure}

Optical coherence tomography (OCT) systems have been implemented so far using free-space or fiber optics components only. Although systems based on fiber optics allow for the compact, inexpensive and portable instrumentation, the final size and price of the devices are still not adequate for many applications. OCT systems composed of many optical components, such as: optical fibers, lenses beam splitters or directional coupler (DC), etc. These components are necessary to relay the optical signal through the OCT system. The combination of all these components makes OCT systems bulky, expensive and complex.

To solve above problem we will fabricate OCT system by using MOEMS fabrication technology. This technology has been proven that the low loss waveguide structure can be easily fabricated with either silicon on insulator waveguide technology or polymer (PMMA, SU8, S1813 etc) based waveguide. Before going for any waveguide fabrication process we need to simulate the system. In this article we simulate polymer based OCT system by using appropriate software codes.

\section{Theory}
Figure \ref{fig1}, shows schematic of a free space OCT system. The heart of the OCT system is a Michelson Interferometer. The low coherence light beam entering at one arm (A1)\ of OCT  is directed to reference (A3)\ and sample (A4)\ arms by beam splitter. The mirror in A3 is moved or scanned to induce a path difference mechanically or optically. The light scattered from the stratified media of the sample beam also return towards the detector. The back scattered light from A4 interfere with the back reflected light from the A3, when the path difference between them is within the coherence 
length of the light source {\cite{Huang,drex}}. OCT measures the interference resulting from different axial locations within the tissue by moving the mirror at the reference arm which generates continuous interference signals along the depth. The depth-resolved sample information from each column of data is appended together and visualized as a 2-D cross-sectional image after signal processing.

We are proposing a polymer based MOEMS OCT system, where Michelson interferometer is fabricated using a waveguide bi-directional coupler. The moving mirror is replaced with an optical delay line generator. The delay line is integrated and
replaced with the mirror component of the MI.
The delay line which can generate temporal bunches
of waves delayed in time is equivalent to the
movement of a mirror.
\begin{figure}
\begin{center}
\includegraphics[width=0.4\columnwidth]{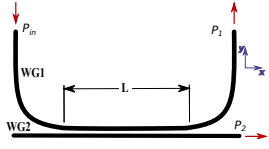}
\end{center}
\caption{\label{fig2}Demonstration of frustrated
total internal reflection. $P_{in}$ may leak through
the waveguide at bottom.}
\end{figure}

\subsection{Bi-directional coupler}
If two similar linear waveguides are parallel and sufficiently closely spaced over a length $L$ 
(coupling length) the energy is transferred from one to another due to the evanescent field
 between them. The fraction of the power coupled is determined by the overlap of the modes 
 in the separate channels. 
In case of the synchronous co-directional coupler (both waveguide channels have same phase 
velocity, ($\Delta k = 0$) all the energy can theoretically be transferred from one waveguide to 
another if  $L=z$ is such that 
\begin{equation}
\kappa z= \frac{(2m+1)\pi}{2},
\end{equation}
with $\kappa$ as the coupling constant and $m = 0, 1, 2...$, Shown in Figure \ref{fig2}.

Generally, $\kappa$ controls the ratio of power
transferred between the waveguides. The value of
$\kappa$ depends on the percentage of overlap of the mode profiles
within the waveguides. The coupling constant is
generally defined as {\cite{Huns}} 
\begin{equation}
\kappa = \frac{2k_x^2 q_x \exp(-q_x c)}{k_z b(q_x^2+k_x^2)}.
\end{equation}
Here, $b$ - channel width, $c$ - separation between the
channels, $k_x$ and $k_z$ are the propagation constants along the two transverse direction ($x$ and $z$ axis) 
to propagation of wave (we suppose to direction of propagation of wave along $y$ axis) and $q_x$ 
is the exponential falloff along the direction of propagation of wave.

This principle is the basis for the operation of directional couplers, by adjusting the suitable
coupling length $L$, any desired portion of energy from a
waveguide can be transferred into adjacent waveguide. 
In our case both waveguide channel have same phase velocity and we need to transfer 
same energy between both sample and reference arm so, we need to choose the coupling length  
such that 50\% of light coupled from one waveguide to the
adjacent one. Accordingly, the condition for 50:50
splitting ratio is \cite{tamir},
\begin{equation}
\frac{L}{2}=\frac{(2m+1)\pi}{4\kappa}.
\end{equation}
\subsection{Optical delay line}

Optical delay line is another important functional component of OCT that provides
 optical delay at the reference arm to get interference signal with the back-scattered 
 optical signals at sample arm and it also attends fast scanning speed and scan depth. 
 Fiber optic delay lines are already been demonstrated \cite{choi}. Compact Silicon-Based Integrated
  Optic time delay line has been fabricated {\cite{yegan,om}} and demonstrated optical time delay. 
  The silicon-on-insulator (SOI) technology is adopted with an incremental time delay of 12.3 ps. 
  Guided- wave optical delay line provide the required precision and are more compact than optical
   fibers \cite{alam}.

\section{Design and Working}
The proposed lab-on a chip device of optical coherence
tomography is exhibited in Figure 3. This design is divided into three sections: P1:~Michelson
interferometer with a 3dB coupler, P2:~Y~coupler for connecting the delay line with 3dB coupler 
and P3:~Optical delay line. 
\begin{figure}
\begin{center}
\includegraphics[width=0.8\columnwidth]{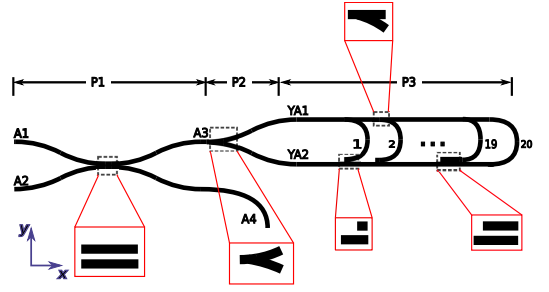}
\end{center}
\caption{\label{fig3}Schematic diagram of all
optical integrated time
domain optical coherence tomography using synchronous directional coupler.
P1-Michelson interferometer, P2-Y-channel light
coupler and decoupler and P3- temporal delay line
mechanism. A1 - A4 : input-output arms of MI, YA1-YA2: Y-channel
couplers. Design of OCT system; Inset: magnification of particular part of system.}
\end{figure}
Further a comparison of Figures 1 and 3, can
be made to understand the arms of A1-A4 of MI
with the proposed design.

\begin{figure}
\begin{center}
\includegraphics[width=0.8\columnwidth]{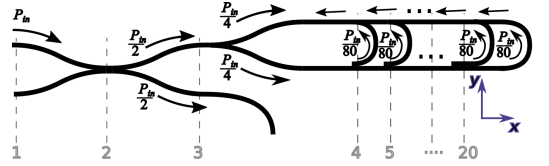}
\end{center}
\caption{\label{fig4} Power distribution requirement
for OCT design. The power distribution is optimized
for best visibility and good modulation index.}
\end{figure}

The power distribution as shown in Figure 4, is
discussed below. As we launched an optical power of $P_{in}$ from A1 arm
of 3dB coupler, the light is divided into two equal parts 
to the arms A3 and A4. The optical power expected in the
arms A3 and A4 is $P_{in}/2$. A4 is sample arm and the
arm A3 is connected to Y coupler. 
Further at P2 region the power is divided to $P_{in}/4$ to each YA1
and YA2 arms. Delay line consist of 20 C-channels separated at
a pre-calculated distance. at YA1 arm, all C-channel 
is directly connected while in YA2 all C-channels (except 20th C-channel) is separated 
by a pre-calculated distances. Gap between lower YA2
arm and the C-channels are calculated for a leakage
of 5\% of optical power only. 

1st C-channel decouples nearly 5\% of optical power to
the YA2 arm, corresponding to an optical power
of $P_{in}/80$. The remaining optical power is passed to  2nd  and
rest of the channels. The design ensure that all
C-channels decouples a only 5\% $P_{in}$, and the remaining
is passed on. While the power is splitted, each
bunch of $P_{in}/80$ optical power also generates a
path delay equivalent to 60$\mu$m with each other.
The path difference induces a temporal delay of
about 0.6ps with next $P_{in}/80$ optical power.
This pulse delay is typically required resolution in
a OCT system. In other words the OCT system can resolve those two layers 
of samples whose optical distance is nearly
30$\mu$m.

\section{Results}
We consider the fabrication of lab-on chip OCT system with
following parameters: \\
Design wavelength ($\lambda$) = 1.55 $\mu$m; 
$n_{core} = 1.48$ (SU 8); $n_{clad} = 1$ (air); \\
Channel width = 1 $\mu$m; Radius of curvature of each C-channel = 6 $\mu$m; \\
The estimated values of coupling length of each C-channel for constant power coupling of $P_{in}/80$
in Table 1.

\begin{table}[h]
\caption{\label{ex}Estimated values of coupling length of each C-channel for constant power coupling of $P_{in}/80$. R:~required percentage coupling for constant power coupling of $P_{in}/80$, $L_{c}$:~ coupling length of particular C channel.}
\begin{center}
\begin{tabular}{ccc||ccc}
C-channel \#& $R$      &$L_{c}$& C-channel  \# & $R$ &    $L_{c}$\\
1&      5.00& 4.0 $\mu$m&11&    10.00&  8.60 $\mu$m\\
2&      5.26&   4.2 $\mu$m&12&  11.11&  9.60 $\mu$m\\
3&      5.56&   4.5 $\mu$m&13&  12.50&  11.2 $\mu$m\\
4&      5.88&   4.8 $\mu$m&14&  14.29&  12.8 $\mu$m\\
5&      6.25&   5.2 $\mu$m&15&  16.67&  15.0 $\mu$m\\
6&      6.67&   5.7 $\mu$m&16&  20.00&  17.4 $\mu$m\\
7&      7.14&   6.2 $\mu$m&17&  25.00&  20.0 $\mu$m\\
8&      7.69&   6.5 $\mu$m&18&  33.33&  27.0 $\mu$m\\
9&      8.33&   7.0 $\mu$m&19&  50.00&  45.0 $\mu$m\\
10& 9.09&       7.7 $\mu$m&20&  100.0&  Directly connected\\
\end{tabular}
\end{center}
\end{table}

Simulation is performed using appropriate software
code, where the complete structure as shown in
Fig. 4 is used. All design parameters discussed
above are incorporated. About 20 grid points as
shown (in dashed gray lines) the figure are the
places where measurements were made. 
\begin{figure}
\begin{center}
\includegraphics[width=0.8\columnwidth]{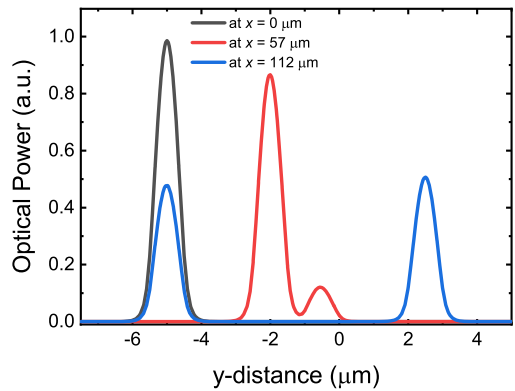}
\end{center}
\caption{\label{fig5} Power output measured from
the P1 as shown in Fig. 3. The optical power is
measured at three cross-sectional locations of
the directional coupler.}
\end{figure}

The Y-coupler as depicted as P1 in the design
as simulated and studied first. The results are
shown in Fig. 5. The measurements made at grid
points 1, 2 and 3 are exhibited. The inset shows
the power profile. The optical power coupled to the
input port A2 is assumed as 100 \% power. At grid
point 3, the power is divided to 50:50 and act as
3dB coupler. 

Simulation of the complete design as given in
Fig. 4 is performed. The results obtained with optimized
values are shown in Figure 6. The optical power profile
values obtained at the grid point locations are
plotted. The light is coupled at A2 port 18-19~$\mu$m,
region, YA1 is located at a distance of 23-24~$\mu$m
region while the C-channels ends at 25-26~$\mu$m
region. YA1 can be located at 31-32~$\mu$m region.

Figure 6a, clearly demonstrates the capability
of all the components P1, P2 and P3. The expected
50:50 ratio may be noted at grid point 3. The
power is further makes round trips via YA1 to
YA2 as well as through YA2 to YA1. Both outputs
reaches the directional coupler and A2 arm. 

The magnified and enhanced view of YA2-C-channel
region is depicted in Fig 6b. The results shows
light coupling between these channels. However,
expected values of 5\% optical power from each
channel could not be found. A exponential decay
of power reduction happens at channels 10 and
beyond. Overall we could see that required delay
is generated, but with slight change in power
ratio. The change power ratio might change the
visibility of the images only. 

\begin{figure}
\begin{center}
\includegraphics[width=0.45\columnwidth]{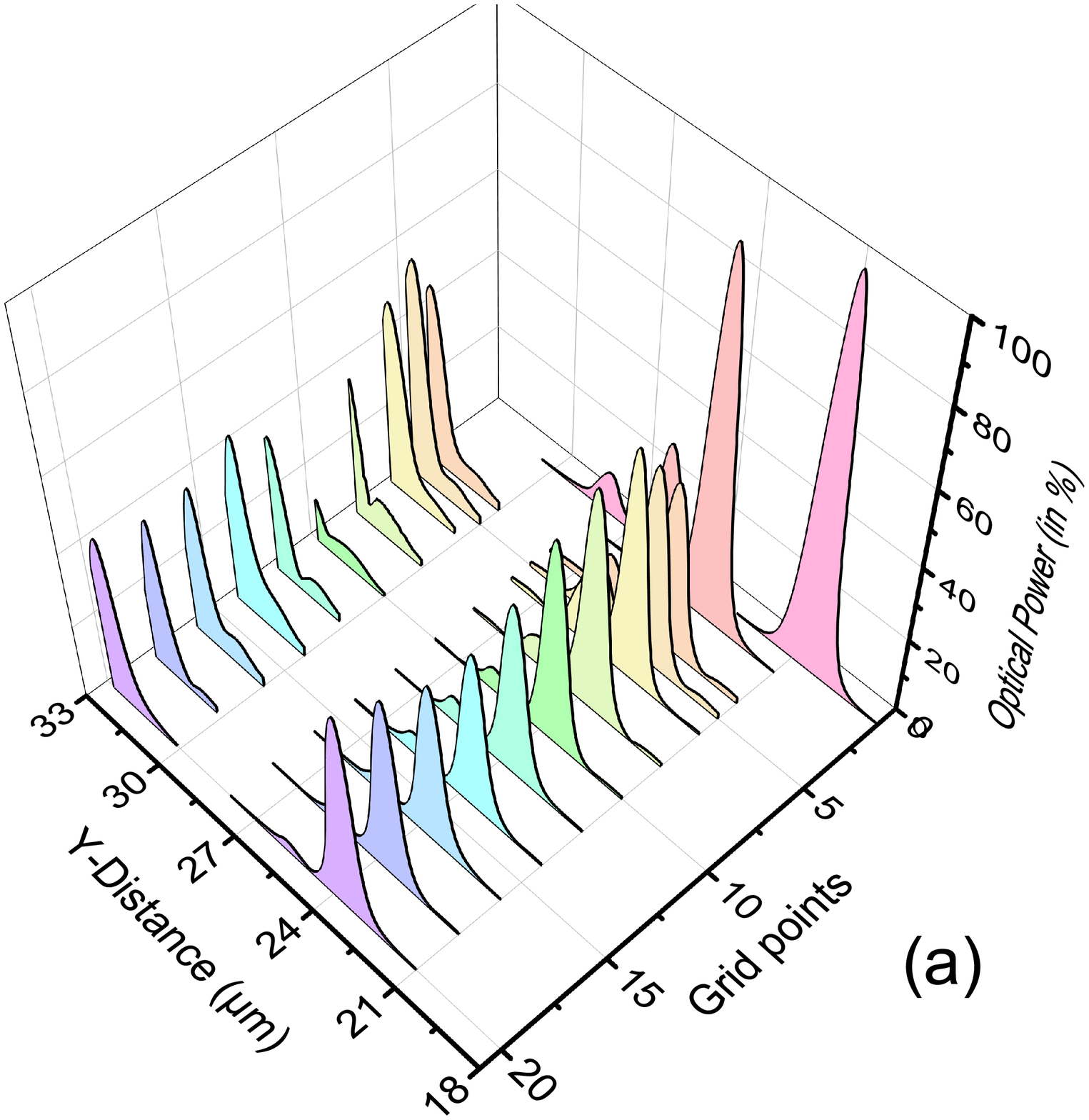}
\includegraphics[width=0.45\columnwidth]{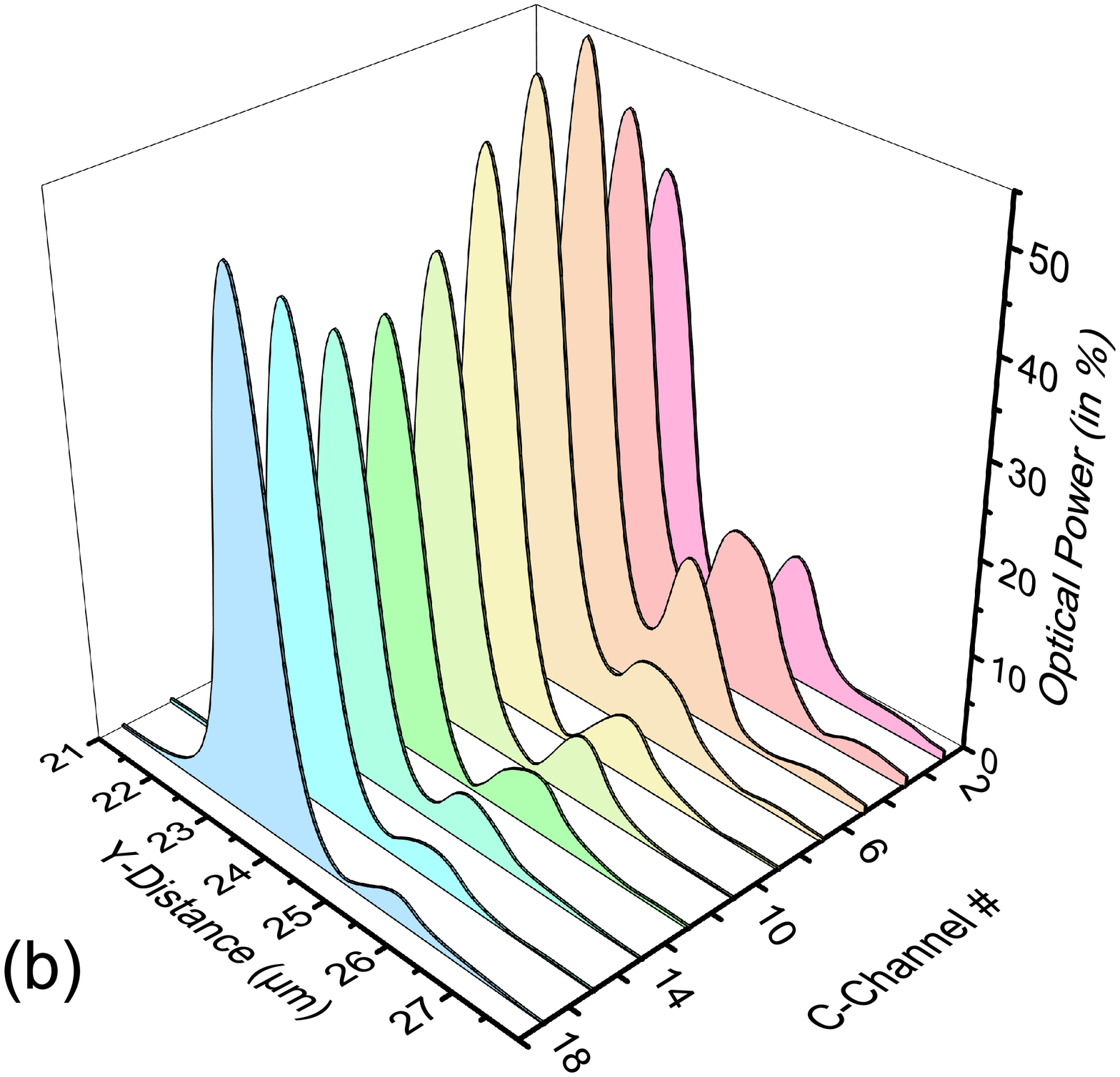}
\end{center}
\caption{\label{fig6} Optical power output as
measured from the complete design as shown in
Fig. 4. (a) Shows all measured locations marked
as 1-20 in Fig. 4. (b) Exhibits the optical power
at small region where the light coupling between the
C-channel and YA2 is made. }
\end{figure}

\section{Conclusions}
 
We proposed a design of all optical optical coherence
tomography system using SU-8 polymer. The simulated
results shows the definite possibility of miniaturized
OCT. The lab-on-a-chip design dimensions are :
length = 1.4mm and width = 50 $\mu$m approx.
The area of the chip after embedding the detector
and light source would be less than 2mm$^2$.
This promises the possibility of fabrication of all optical 
OCT for point of contact devices as well as for
embedded OCT chip a reality. 

\section*{Acknowledgments}
ASB and JTA acknowledge the financial support
received from SERB, New Delhi GoI, under the project
number CRG/2018/003871.

\end{document}